\documentclass[journal]{IEEEtran}
\usepackage{amsmath,graphicx,url,times,hyperref}
\usepackage[usenames, dvipsnames]{color}
\usepackage{multirow}
\ifCLASSINFOpdf
\else
\fi
\begin{document}
\title{Unsupervised Feature Learning Based on Deep Models for Environmental Audio Tagging}

\author{Yong~Xu,~
	    Qiang~Huang,~
        Wenwu~Wang,~\IEEEmembership{Senior Member,~IEEE,}
        Peter~Foster,~
        Siddharth~Sigtia,~
        Philip~J.~B.~Jackson,~
        and~Mark~D.~Plumbley,~\IEEEmembership{Fellow,~IEEE}% <-this % stops a space
%\thanks{M. Shell was with the Department
%of Electrical and Computer Engineering, Georgia Institute of Technology, Atlanta,
%\thanks{Manuscript received July 12, 2016; revised October 26, 2016.}
\thanks{Yong Xu, Qiang Huang, Wenwu Wang, Philip~J.~B.~Jackson and Mark~D.~Plumbley are with the Centre for Vision, Speech and Signal Processing, University of Surrey, GU2 7XH, UK. Peter~Foster and Siddharth~Sigtia are with the School of Electronic Engineering and Computer Science, Queen Mary University of London, London, E1 4NS, UK. (Emails: yx0001@surrey.ac.uk, q.huang@surrey.ac.uk, w.wang@surrey.ac.uk, p.a.foster@qmul.ac.uk, s.s.sigtia@qmul.ac.uk, p.jackson@surrey.ac.uk, m.plumbley@surrey.ac.uk)}% <-this % stops a space
\thanks{This work was supported by the Engineering and Physical Sciences Research Council (EPSRC) of the UK under the grant EP/N014111/1.}}

\maketitle

\begin{abstract}
\textcolor{red}{Environmental audio tagging
aims to predict only the presence or absence of certain acoustic events in the interested acoustic scene.
In this paper we make contributions to audio tagging in two parts, respectively, acoustic modeling and feature learning. We propose to use a shrinking deep neural network (DNN) framework incorporating unsupervised feature learning to handle the multi-label classification task. For the acoustic modeling, a large set of contextual frames of the chunk are fed into the DNN to perform a multi-label classification for the expected tags, considering that only chunk (or utterance) level rather than frame-level labels are available. Dropout and background noise aware training are also adopted to improve the generalization capability of the DNNs.
For the unsupervised feature learning, we propose to use a symmetric or asymmetric deep de-noising auto-encoder (sDAE or aDAE) to generate new data-driven features from the Mel-Filter Banks (MFBs) features. The new features, which are smoothed against background noise and more compact with contextual information, can further improve the performance of the DNN baseline.
Compared with the standard Gaussian Mixture Model (GMM) baseline of the DCASE 2016 audio tagging challenge, our proposed method obtains a significant equal error rate (EER) reduction from 0.21 to 0.13 on the development set. The proposed aDAE system can get a relative 6.7\% EER reduction compared with the strong DNN baseline on the development set. Finally, the results also show that our approach obtains the state-of-the-art performance with 0.15 EER on the evaluation set of the DCASE 2016 audio tagging task while EER of the first prize of this challenge is 0.17.}
\end{abstract}

\begin{IEEEkeywords}
Environmental audio tagging, deep neural networks, unsupervised feature learning, deep denoising auto-encoder, DCASE 2016.
\end{IEEEkeywords}

% \fi
\IEEEpeerreviewmaketitle

\section{Introduction}
\label{sec:intro}
\IEEEPARstart{A}{s} smart mobile devices are widely used in recent years, huge amounts of
multimedia recordings are generated and uploaded to the web every day. 
These recordings, such as music, field sounds, broadcast news, and
television shows, contain sounds from a wide variety of sources.
The demand for analyzing these sounds is increasing, e.g. for
automatic audio tagging \cite{Alberti2009}, 
audio segmentation \cite{foote2000automatic} and 
audio context classification \cite{Allegro01automaticsound, Picart2015}.
%%%%%%%%%%%%%%%%%%%%%%%%%%%%%%%%%%%%%%%(
Due to the technology and customer need, there have been some applications of audio processing
in different scenarios,
such as urban monitoring, surveillance, health care, music retrieval, and customer video. 

%In urban monitoring, the audio scene analysis, as an effective complement to video scene recognition,
%can provide road traffic estimation, and road traffic event detection and localization \cite{Chen1997}.
%Relying on relative cheap and scalable sensors, Gubbi \textit{et al}. developed a system to
%present visualization of real-time data via the internet \cite{Gubbi2013}.
%In surveillance of suspicious events accompanied by
%banging sounds and screaming,
%audio analysis can take us closer to semantics than video 
%analysis and is also computationally more efficient\cite{Radhakrishnan2005}.
%For health care, breath sounds have been shown very valuable for 
%diagnosis of obstructive sleep apnea \cite{Alshaer2014,Brian2015}.
%Content based music information retrieval
%%!!!!!!!!,as a successful application of audio signal processing, 
%can now well process different queries by searching for the
%relevant contents in a million song database \cite{Bertin2011}. 
%For consumer video analysis,
%Zhang \textit{et al}. \cite{Zhang2001} proposed a system
%to segment and classify audio from movies or TV programs
%into multiple classes such as speech, music and environmental
%sound.  

\textcolor{red}{For environmental audio tagging, there is a large amount of audio data online, e.g. from Youtube or Freesound, which are labeled with tags. How to utilize them, predict them and further add some new tags on the related audio is a challenge. The environmental audio recordings are more complicated than the pure speech or music recordings due to the multiple acoustic sources and incidental background noise. This will make the acoustic modeling more difficult. On the other hand, one acoustic event (or one tag) in environmental audio recordings might occur in several long temporal segments. A compact representation of the contextual information will be desirable in the feature domain.}
%Therefore, the focus of this work will be on these two issues.}

In traditional methods, a common approach is to convert low-level acoustic features into
``bag of audio words'' \cite{chen2010,riley2008, shao2004, cai2005,Sainath2007}.
%The second type of techniques is based on only weakly labeled data \cite{DBLP:journals/corr/0003R16}, e.g. 
%audio tagging. 
%In the first type of methods,
K-means, as an unsupervised clustering method, has been
widely used in audio analysis \cite{chen2010} and music retrieval \cite{riley2008, shao2004}.
In \cite{cai2005}, Cai \textit{et al}. replaced K-means with a spectral clustering-based scheme 
to segment and cluster the input stream into audio elements.
Sainath \textit{at al.} \cite{Sainath2007} derived an audio segmentation method
using Extended Baum-Welch (EBW) transformations for estimating parameters of Gaussian mixtures.
Shao \textit{at al.} \cite{shao2004} proposed to use a measure of similarity derived
by hidden Markov models to cluster segment of audio streams. 
Xia \textit{et al.} \cite{xia2011} used Eigenmusic and Adaboost to separate rehearsal
recordings into segments, and an alignment process to organize segments.
Gaussian mixture model (GMM), as a common model, was also used as the official baseline method in DCASE 2016 for audio tagging \cite{dcase_t4}. Recently, in \cite{DBLP:journals/corr/0003R16}, a Support Vector Machine (SVM) based Multiple Instance Learning (MIL) system was also presented for audio tagging and event detection. The details of the GMM and SVM methods are presented in the appendix of this paper. However, these methods can not well utilize the contextual information and the potential relationship among different event classes.

%These applications have motivated the development of 
%new audio processing methods, but also introduced new challenges, such as the lack of accurately annotated data and acoustic feature learning, the two problems to be discussed in this paper. 
%In many previous methods, in order to train a model, annotated data are often required, where the desirable audio events need to be clearly labeled. Although supervised approaches have been widely used, 
%their effectiveness relies heavily on the
%quantity and quality of the training data. 
%Nowdays, it is easy to collect a large amount of audio data
%online, e.g. from Youtube or Freesound, however most of them are either not
%labeled at all or only labeled very weakly using a small amount of metadata
%or simple conceptual tagging.     
%It is clear that processing such audio data in a
%supervised manner will be quite hard because 
%manually labeling data is very time-consuming. 
%Moreover, the quality of audio data is often unsatisfying
%because of the variations in recording devices used and the environment where
%the audio data are recorded.  
%Accordingly, learning robust audio features to handle 
%acoustic variation becomes highly desirable.    

%Deep learning technologies have obtained great successes in speech, 
%image and video fields \cite{xu2014experimental, xu2015regression, hinton2012deep, krizhevsky2012imagenet} 
%since Hinton and Salakhutdinov showed the insights 
%using a greedy layer-wise unsupervised learning procedure 
%to train a deep model in 2006 \cite{hinton2006reducing}. 
The deep learning methods were also investigated for related tasks,
like acoustic scene classification \cite{petetin2015deep}, 
acoustic event detection \cite{cakir2015polyphonic} and unsupervised feature learning \cite{Hamel2009} and better performance could be obtained in these tasks. 
For music tagging task, \cite{dieleman2014end, choi2016automatic} 
have also demonstrated the superiority of deep learning methods. 
Recently, the deep learning based methods have also been widely used for environmental audio tagging \cite{lidy2016cqt, cakirdomestic}, a newly proposed task in DCASE 2016 challenge \cite{dcase_t4} based on the CHiME-home dataset \cite{foster2015}. However, it is still not clear what would be appropriate input features, objective functions and the model structures for deep learning based audio tagging. Furthermore, only the chunk-level instead of frame-level labels are available in the audio tagging task. 
Multiple acoustic events could occur simultaneously with interfering background noise, 
for example, \textit{child speech} could exist with \textit{TV sound} for several seconds. Hence, a more robust deep learning method is needed to improve the audio tagging performance.

Deep learning was also widely explored in feature learning \cite{coates2010analysis, bengio2012unsupervised}. These works have demonstrated that data-driven learned features can get better performance than the expert-designed features.
%Neural network based bottleneck feature \cite{grezl2007probabilistic} in speech recognition task is one successful type of supervised learned feature where the bottleneck feature is extracted from the middle layer of a DNN classifier. 
In \cite{coates2010analysis}, four unsupervised learning algorithms, K-means clustering, restricted Boltzmann machine (RBM), Gaussian mixtures and auto-encoder are explored in image classification. Compared with RBM, auto-encoder is a non-probabilistic feature learning paradigm \cite{bengio2012unsupervised}. For the audio tagging task, Mel-frequency Cepstral Coefficients (MFCCs) and Mel-Filter Banks (MBKs) are commonly adopted as the basic features. However it is not clear whether they are the best choice for audio tagging.

\textcolor{red}{In this paper, we propose a robust deep learning framework for the audio tagging task, with focuses mainly on the following two parts, acoustic modeling and unsupervised feature learning, respectively. For the acoustic modeling, we investigate deep models with shrinking structure, which can be used to reduce the model size, accelerate the training and test process \cite{zhang2014improving}.
Dropout \cite{dahl2013improving} and background noise aware training \cite{xu2014dynamic} are also adopted to further improve the tagging performance in the DNN-based framework. Different loss functions and different basic features will be also compared for the environmental audio tagging task.
%The neural network structure was also successfully used in 
%image segmentation \cite{long2015fully}.
For the feature learning, we propose a symmetric or asymmetric deep de-noising auto-encoder (sDAE or aDAE) based unsupervised method to generate a new feature from the basic features. There are two motivations here, the first is the background noise in the environmental audio recordings which will introduce some mismatch between the training set and the test set. However, the new feature learned by the DAE can mitigate the impact of background noise. The second motivation is that compact representation of the contextual frames is needed for the reason that only chunk-level labels are available. The proposed sDAE or aDAE can encode the contextual frames into a compact code, which can be used to train a better classifier.}

%\textcolor{red}{The contribution of this paper is three-fold. First, we proposed a robust DNN-based framework for the newly proposed environmental audio tagging task in DCASE 2016 challenge. Then different loss functions, namely mean squared error (MSE) and binary cross-entropy (or called logistic regression), and different basic features, namely Mel-frequency cepstral coefficients (MFCCs) and Mel-filter banks (MBKs), were compared for the environmental audio tagging task. Finally, we propose a DAE based unsupervised feature learning to smooth the background noise and generate a compact representation of the contextual frames. Experimental results show that we get the state-of-the-art performance with 14.8\% EER on the evaluation set of the DCASE2016 audio tagging challenge.}

%To get a better prediction of the tags, a deep pyramid structure 
%is designed with gradually shrinked size of layers. 
%This deep pyramid structure can reduce the non-correlated interferences 
%in the whole audio features while focusing on extracting the robust 
%high-level features related to the target tags. 

%%%%%%%%%%%%%%%%%%%%%%%%%%%%%%%%%%%%%%%%%%%%%%%%%%%(
%Furthermore, to learn the robust features against acoustic variations
%in the training and test data, we will also apply DNN based autoencoder
%to the MFCCs. 
%%%%%%%%%%%%%%%%%%%%%%%%%%%%%%%%%%%%%%%%%%%%%%%%%%%)

The rest of the paper is organized as follows.
%In section \ref{sec:related_work}, we will introduce the related traditional methods and their drawbacks. 
We present our robust DNN-based framework in section \ref{sec:fully_dnn}. The proposed deep DAE-based unsupervised feature learning will be presented in section \ref{sec:dae_FeaLearn}. The data description and experimental setup
will be given in section \ref{sec:exp}. We will show
the related results and discussions in section \ref{sec:results}, and finally draw a conclusion in section \ref{sec:conclusions}. Appendix will introduce the GMM and SVM based methods in detail, which will be used as baselines for performance comparison in our study.

\section{Robust DNN-based audio tagging}
\label{sec:fully_dnn}
DNN is a non-linear multi-layer model for extracting robust features related to a specific classification \cite{hinton2012deep} or regression \cite{xu2015regression} task. The objective of the audio tagging task is to perform multi-label classification on audio chunks (i.e. assign one or more labels to each audio chunk of a length e.g. four seconds in our experiments). The labels are only available for chunks, but not frames. Multiple events may happen at many particular frames.
%Hence, the common frame-level cross entropy based loss function cannot be adopted. We propose a method to encode the whole or almost whole chunk.

\subsection{DNN-based multi-label classification}
\label{ssec:fully_dnn}
\begin{figure}[t]
	\centering
	\centerline{\includegraphics[width=\columnwidth]{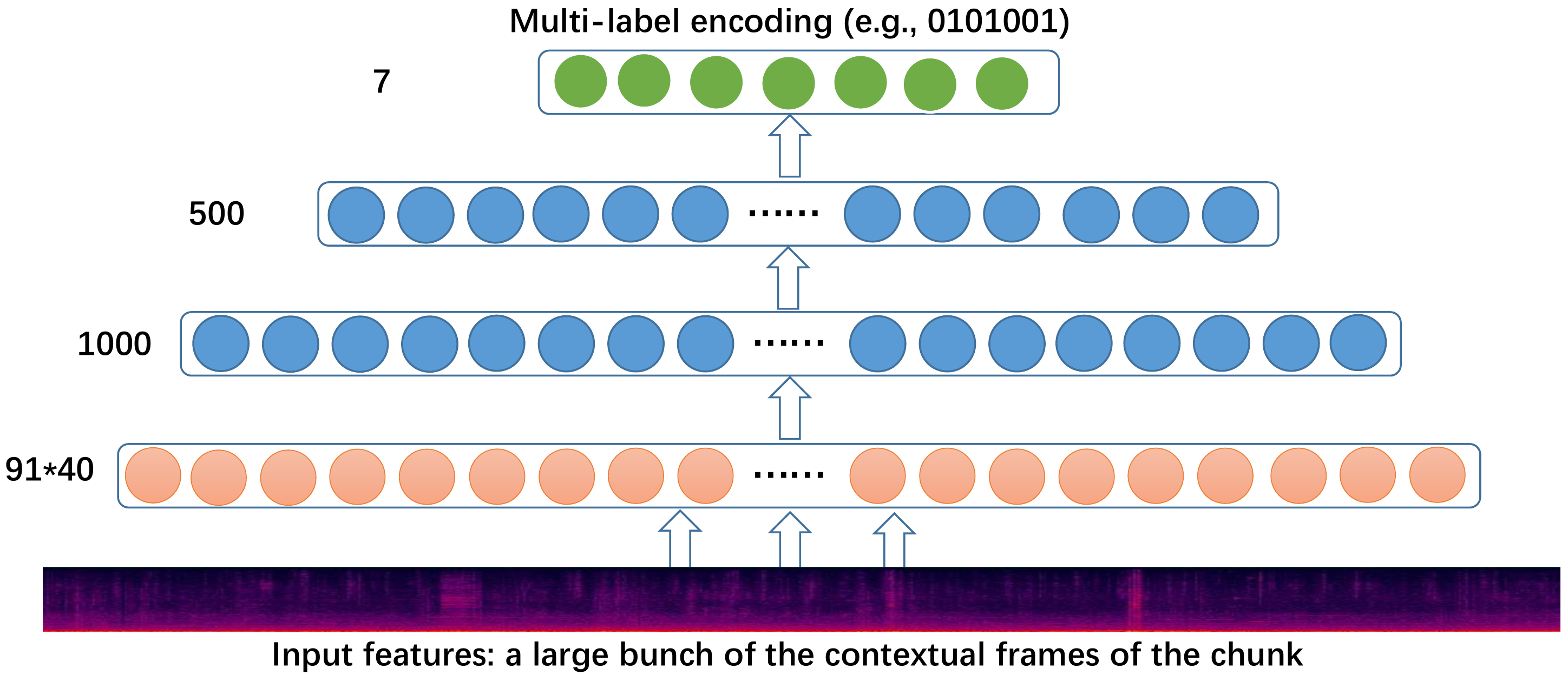}}
	\caption{DNN-based environmental audio tagging framework using the shrinking structure \cite{zhang2014improving}.}
	\label{fig:dnn_tag}
\end{figure}

\textcolor{red}{Fig. \ref{fig:dnn_tag} shows the proposed DNN-based audio tagging framework using the shrinking structure, i.e., the hidden layer size is gradually reduced through depth. In \cite{zhang2014improving}, it is shown that this structure can reduce the model size, training and test time without losing classification accuracy. Furthermore, this structure can serve as a deep PCA \cite{hinton2006reducing} to reduce the redundancy and background noise in the audio recordings.}
With the proposed framework, a large set of features of the chunk are encoded into a vector with values $\{0,1\}$. Sigmoid was used as the activation function of the output layer to learn the presence probability of certain events. Rectified linear unit (ReLU) is the activation function for hidden units. \textcolor{red}{Mean squared error (MSE) and binary cross-entropy were adopted and compared as the objective function. As the labels of the audio tagging are binary values, binary cross-entropy can get a faster training and better performance than MSE \cite{zhou1998learning}. A stochastic gradient descent algorithm is performed in mini-batches with multiple epochs to improve learning convergence as follows,}
\begin{equation}
E_{mse}=\frac{1}{N}\sum_{n=1}^{N}\|\hat{\textbf{T}}_{n} (\textbf{X}{_{n-\tau}^{n+\tau}},\textbf{W}, \textbf{b})-\textbf{T}_{n}\|_{2}^2
\label{eq:DNNerrors_mse}
\end{equation}
\begin{equation}
E_{bce}=-\sum_{n=1}^{N}\|\textbf{T}_{n}\text{log}\hat{\textbf{T}}_{n}+(1-\textbf{T}_{n})\text{log}(1-\hat{\textbf{T}}_n)\|
\label{eq:DNNerrors_bce}
\end{equation}
\begin{equation}
\hat{\textbf{T}}_{n}=(1+\text{exp}(-\textbf{O}))^{-1}
\label{eq:DNN_hidden_sigmoid}
\end{equation}
where $E_{mse}$ and $E_{bce}$ are the mean squared error and binary cross-entropy, $\hat{\textbf{T}}_{n}(\textbf{X}{_{n-\tau}^{n+\tau}},\textbf{W},\textbf{b})$ and $\textbf{T}_{n}$ denote the estimated and reference tag vector at sample index $n$, respectively, with $N$ representing the mini-batch size, $ \textbf{X}{_{n-\tau}^{n+\tau}} $ being the input audio feature vector where the window size of context is $2\tau+1$.
It should be noted that the input window size should cover a large set of contextual frames of the chunk considering the fact that the reference tags are in chunk-level rather than frame-level.
%However, slightly relaxing the window size without covering all of the chunk frames could increase the total number of training samples for DNN. It can improve the performance as observed in our experiments.
The weight and bias parameters to be learned are denoted as $(\textbf{W}, \textbf{b})$.
The DNN linear output is defined as $\textbf{O}$ before the Sigmoid activation function is applied.

The updated estimate of $ \textbf{W}^\ell $ and $ \textbf{b}^\ell $ in the $\ell$-th layer, with a learning rate $\lambda$, can be computed iteratively as follows:
\begin{equation}
(\textbf{W}^\ell, \textbf{b}^\ell)~\leftarrow~(\textbf{W}^\ell, \textbf{b}^\ell)-\lambda\frac{\partial{E}}{\partial{(\textbf{W}^\ell, \textbf{b}^\ell)}},~1\leq{\ell}\leq{L+1}
\end{equation}
where $L$ denotes the total number of hidden layers and the ($L+1$)-th layer represents the output layer.

During the learning process, the DNN can be regarded as an encoding function, and the audio tags are automatically predicted.
%Hence the multi-label regression rather than classification can be conducted. 
The background noise may exist in the audio recordings which may lead to mismatch between the training set and the test set. To address this issue, two additional methods are given below to improve the generalization capability of DNN-based audio tagging. Alternative input features, eg., MFCC and MBK features, are also compared.

\subsection{Dropout for the over-fitting problem}
\label{ssec:dropout}
Deep learning architectures have a natural tendency towards over-fitting especially when there is little training data. This audio tagging task only has about four hours training data with imbalanced training data distribution for each type of tag, e.g., much fewer samples for event class `b' compared with other event classes in the DCASE 2016 audio tagging task. Dropout is a simple but effective way to alleviate this problem \cite{dahl2013improving}. In each training iteration, the feature value of every input unit and the activation of every hidden unit are randomly removed with a predefined probability (e.g., $ \rho $). These random perturbations effectively prevent the DNN from learning spurious dependencies. At the decoding stage, the DNN scales all of the weights involved in the dropout training by $(1-\rho)$, regarded as a model averaging process \cite{hinton2012improving}.

%A mismatch problem may also exist in this task, and the audio segments in the testing set could be totally different from the existing audio segments in the training set due to the presence of background noise. Thus Dropout should be adopted to improve its robustness to generalize to the variation in the segments from the test set.

\subsection{Background noise aware training}
Different types of background noise in different recording environments could lead to the mismatch problem between the testing chunks and the training chunks. To alleviate this, we propose a simple background noise aware training (or adaptation) method. To enable this noise awareness, the DNN is fed with the primary audio features augmented with an estimate of the background noise. In this way, the DNN can use additional on-line information of background noise to better predict the expected tags. The background noise is estimated as follows:
\begin{equation}
\textbf{V}_{n}=[\textbf{Y}_{n-\tau},...,\textbf{Y}_{n-1},\textbf{Y}_{n},\textbf{Y}_{n+1},...,\textbf{Y}_{n+\tau}, \hat{\textbf{Z}}_{n}]
\end{equation}
\begin{equation}
\hat{\textbf{Z}}_{n}=\dfrac{1}{T}\sum_{t=1}^{T}\textbf{Y}_{t}
\label{eq:nat}
\end{equation}
where the background noise $\hat{\textbf{Z}}_{n}$ is fixed over the utterance and estimated using the first $T$ frames. Although this noise estimator is simple, a similar idea was shown to be effective in DNN-based speech enhancement \cite{xu2015regression, xu2014dynamic}.

\subsection{Alternative input features for audio tagging}
\textcolor{red}{Mel-frequency Cepstral Coefficients (MFCCs)
have been used in environmental sound source classification \cite{Radhakrishnan2005,Cai2006},
however, some previous work \cite{Hamel2010, Cotton2011} showed
that the use of MFCCs is not the best choice 
as they are sensitive to background noise. Mel-filter bank (MBK) features have already been demonstrated to be better than MFCCs in speech recognition in the DNN framework \cite{seltzer2013investigation}. However it is not clear whether this is also the case for the audio tagging task using DNN models. 
Recent studies in audio classification have also shown that 
accuracy can be boosted by using
features that are learned in an unsupervised
manner, with examples in the areas of bioacoustics \cite{Stowell2014} and music
\cite{Vaizman2014}. We will study the potential of such methods for audio tagging and present a DAE-based feature learning method in following section.}

%In \cite{Stowell2014, Dieleman2013}, Spherical k-means \cite{Coates2012} works
%as a more effective feature learning method than the original K-means
%by using unit L2 norm as a constraint on the centroids.
%To further improve feature learning, some pre-processing methods, e.g. principal component analysis (PCA) whitening \cite{Coates2012}
%by decorrelating the input dimensions and post-processing methods, e.g.
%pooling the features across wide windows \cite{Hamel2010},
%are also utilized.

\begin{figure}[t]
	\centering
	\centerline{\includegraphics[scale=0.4]{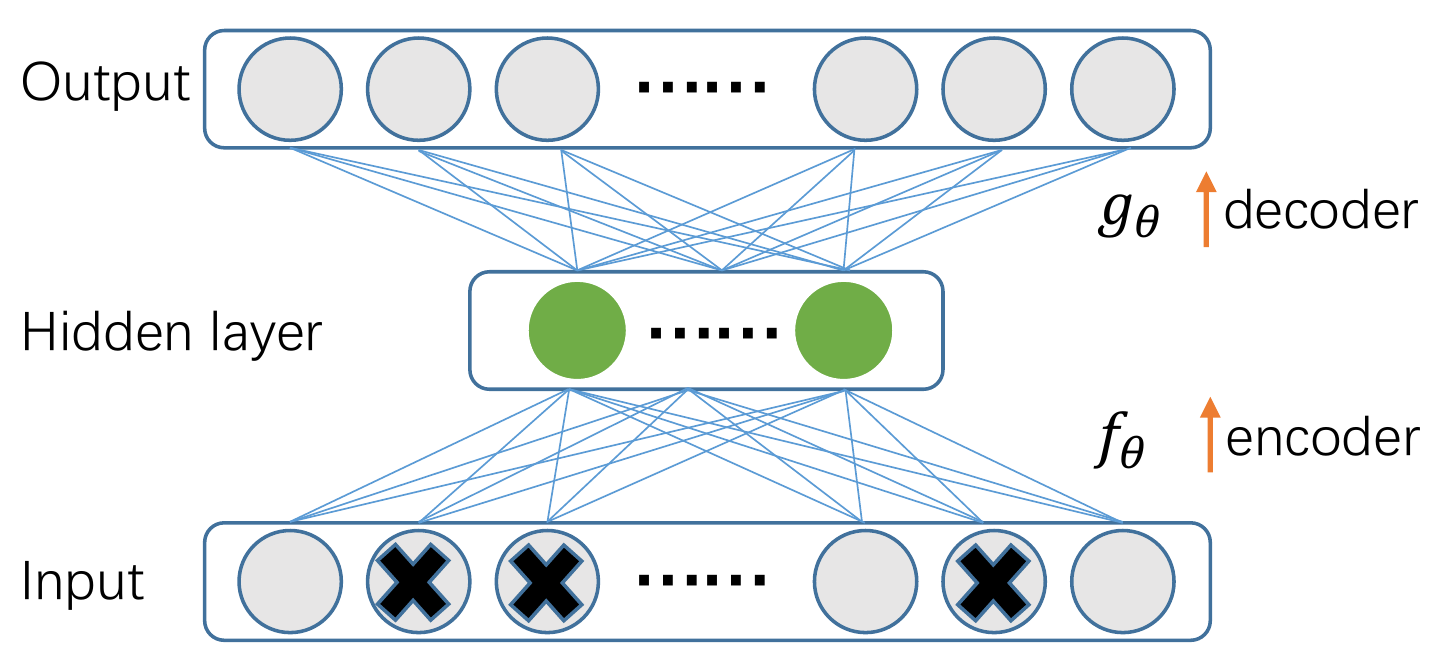}}
	\caption{A typical one hidden layer of de-noising auto-encoder \cite{vincent2008extracting} structure with an encoder and a decoder. Some input units are set to zero by the Dropout process (shown by a black cross ``X") to train a more robust system.}
	\label{fig:autoencoder}
\end{figure}
\section{Proposed Deep asymmetric DAE based unsupervised feature learning}
\label{sec:dae_FeaLearn}
MFCCs and MBKs are used as the basic features for the training of DNN-based predictor in this work. MFCCs and MBKs are well-designed features derived by experts based on the human auditory perception mechanism \cite{molau2001computing}. Recently, more supervised or unsupervised feature learning works have demonstrated that data-driven learned features can offer better performance than the expert-designed features. Neural network based bottleneck feature \cite{grezl2007probabilistic} in speech recognition is one such type of feature, extracted from the middle layer of a DNN classifier. Significant improvement can be obtained after it is fed into a subsequent GMM-HMM (Hidden Markov Model) system and compared with the basic features. However, for the audio tagging task, the tags are weakly labeled and not accurate through the multiple voting scheme. Furthermore, there are lots of related audio files without labels on the web. Hence to use these unlabeled data, we proposed a DAE based unsupervised feature learning method.

\textcolor{red}{Specifically, for environmental audio tagging task, disordered background noise exists in the recordings which may lead to the mismatch between the training set and the test set. DAE-based method can mitigate the effect of background noise and focus on more meaningful acoustic event patterns. Another motivation is that the compact representation of the contextual frames is needed since the labels are in chunk-level rather than frame-level.}
	 
\begin{figure*}[t]
	\centering
	\centerline{\includegraphics[scale=0.4]{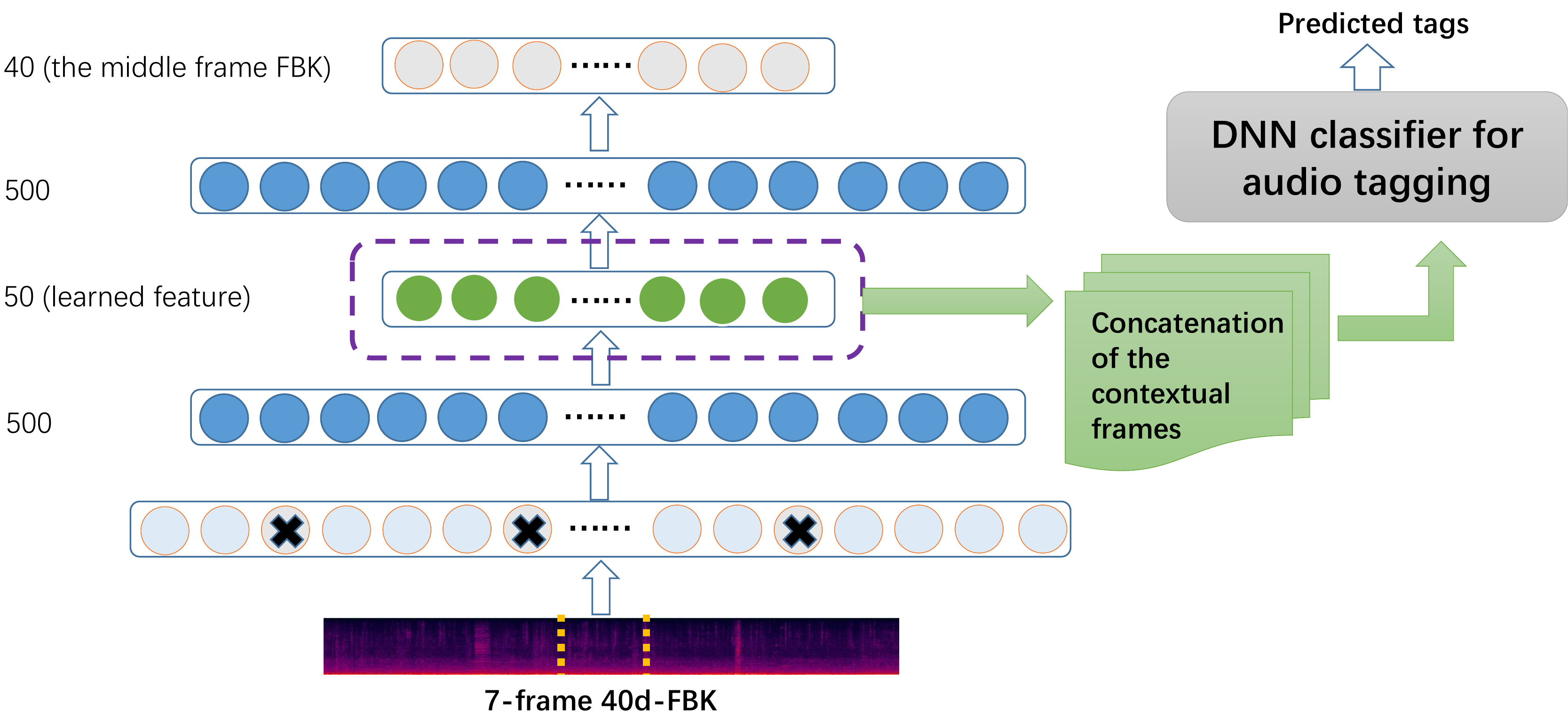}}
	\caption{The framework of deep asymmetric DAE (aDAE) based unsupervised feature learning for audio tagging. The first three layers are the encoder while the last three layers are the decoder. The weights between the encoder and the decoder are untied to retain more contextual information into the bottleneck layer (shown in the dashed rectangle).}
	\label{fig:dae_fl}
\end{figure*}
An unsupervised feature learning algorithm is used to discover features from the unlabeled data. For this purpose, the unsupervised feature learning algorithm takes the dataset $X$ as input and outputs a new feature vector. In \cite{coates2010analysis}, four unsupervised learning algorithms, K-means clustering, restricted Boltzmann machine (RBM), Gaussian mixtures and auto-encoder are explored in image classification. Among them, RBM and auto-encoder are widely adopted to get new features or pretrain a deep model. Compared with RBM, auto-encoder is a non-probabilistic feature learning paradigm \cite{bengio2012unsupervised}. The auto-encoder explicitly defines a feature extracting function, called the \textit{encoder}, in a specific parameterized closed form. It also has another closed-form parametrized function, called the \textit{decoder}. The encoder and decoder function are denoted as $f_{\theta}$ and $g_{\theta}$, respectively. \textcolor{red}{To force the hidden layers to discover more robust features, the de-noising auto-encoder \cite{vincent2008extracting} introduces a stochastic corruption process applied to the input layer. Dropout \cite{dahl2013improving} is used here to corrupt the input units. Compared with the auto-encoder, DAE can discover more robust features and prevent it from simply learning the identity function \cite{vincent2008extracting}.} Fig. \ref{fig:autoencoder} shows a typical one hidden layer of DAE structure with an encoder and a decoder. The encoder generates a new feature vector $\textbf{h}$ from an input $\textbf{x}={x^{(1)},...,x^{(T)}}$. It is defined as, 
\begin{equation}
\textbf{h}=f_{\theta}({\tilde{\textbf{x}}})=s_f(\textbf{W}\tilde{\textbf{x}}+\textbf{b})
\label{eq:ae_encoder}
\end{equation}
where $\textbf{h}$ is the new feature vector or new representation or code \cite{bengio2012unsupervised} of the input data $\textbf{x}$ with the corrupted version $\tilde{\textbf{x}}$. $s_f$ is the non-linear activation function. $\textbf{W}$ and $\textbf{b}$ denote the weights and bias of the encoder, respectively. On the other hand, the decoder, $g_{\theta}$ can map the new representation back to the original feature space, namely producing a reconstruction ${\hat{\textbf{x}}}=g_{\theta}(\textbf{h})$.
\begin{equation}
\hat{\textbf{x}}=g_{\theta}(\textbf{h})=s_g(\textbf{W}'\textbf{h}+\textbf{b}')
\label{eq:ae_decoder}
\end{equation}
where ${\hat{\textbf{x}}}$ is the reconstructed feature which is the approximation of the input feature. $s_g$ is the non-linear activation function of the decoder. $\textbf{W}'$ and $\textbf{b}'$ denote the weights and bias of the decoder. Here $\textbf{W}$ and $\textbf{W}'$ are not tied, namely $\textbf{W}' \neq \textbf{W}^T$.
The set of parameters $\theta=\{\textbf{W},\textbf{b},\textbf{W}',\textbf{b}'\}$ of the auto-encoder are updated to incur the lowest reconstruction error $L(\textbf{x},{\hat{\textbf{x}}})$, which is a measure of the distance between the input $\textbf{x}$ and the output. ${\hat{\textbf{x}}}$. The general loss function for the de-noising auto-encoder \cite{vincent2008extracting} training can be defined as,
\begin{equation}
\varGamma_{AE}(\theta)=\sum_{t}L(x^{(t)},g_{\theta}(f_{\theta}(\tilde{x}^{(t)})))
\label{eq:ae_lossfunction}
\end{equation}
Furthermore, the DAE can be stacked to obtain a deep DAE. The DAE is actually an advanced PCA with the non-linear activation functions \cite{hinton2006reducing}.

\begin{figure}[t]
	\centering
	\centerline{\includegraphics[width=\columnwidth]{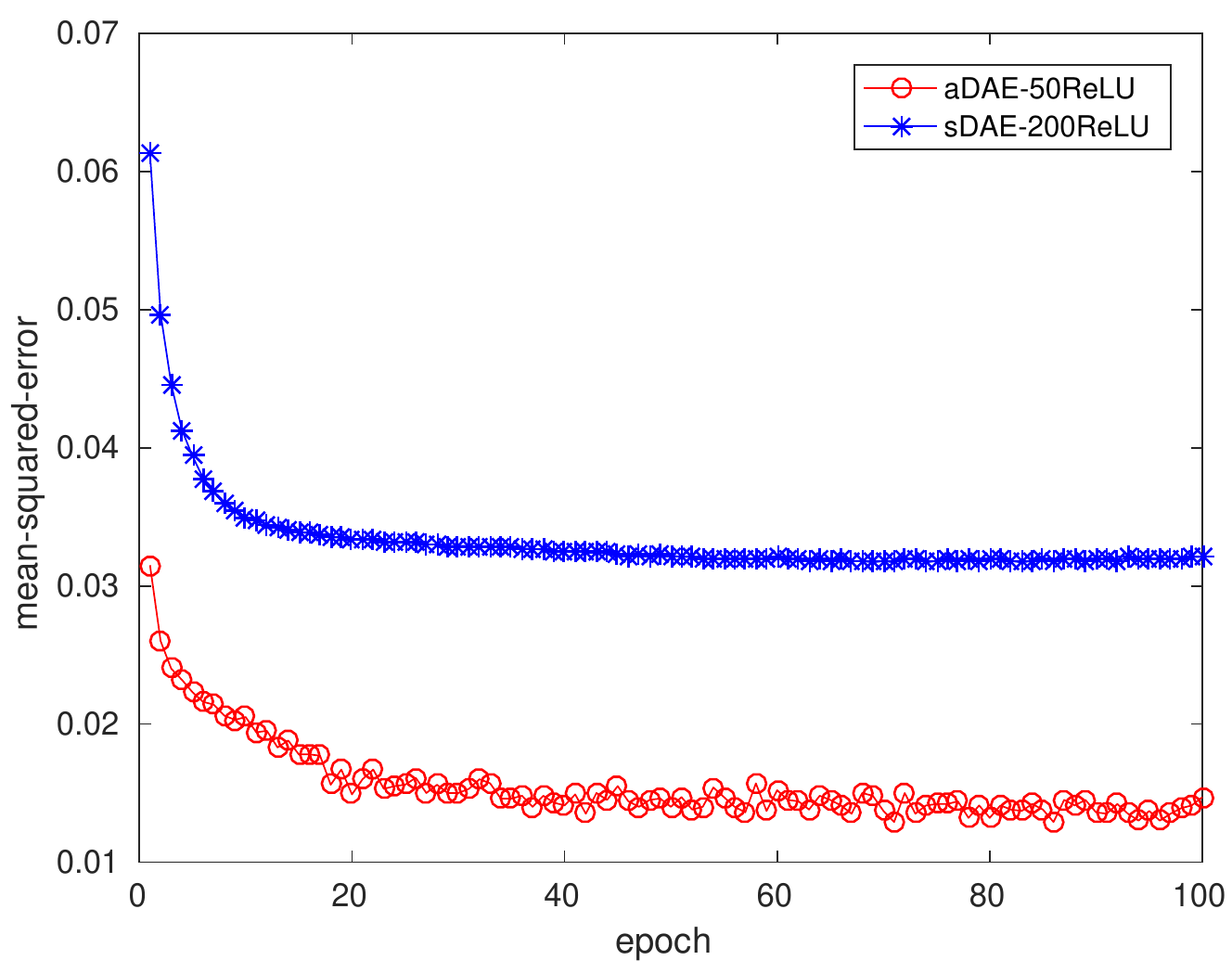}}
	\caption{The reconstruction error over the CV set of the asymmetric DAE with 50 ReLU units in the bottleneck layer (denoted as aDAE-50ReLU) and the symmetric DAE with 200 ReLU units in the bottleneck layer (denoted as sDAE-200ReLU). }
	\label{fig:cv_linear_relu_ser}
\end{figure}

In Fig. \ref{fig:dae_fl}, the framework of deep asymmetric DAE (aDAE) based unsupervised feature learning for audio tagging is presented. It is a deep DAE stacked by simple DAE with random initialization. To utilize the contextual information, multiple frames MBK features are fed into the deep DAE. A typical DAE is a symmetric structure (sDAE) with the same size as the input. However here the deep DAE is only designed to predict the middle frame feature. This is because the more predictions in the output means the more memory needed in the bottleneck layer. In our practice, the deep DAE would generate a larger reconstruction error if multiple frames features were designed as the output with a narrow bottleneck layer. This leads to an inaccurate representation of the original feature in a new space. Nonetheless, with only the middle frame features in the output, the reconstruction error is smaller. Fig. \ref{fig:cv_linear_relu_ser} plots the reconstruction error between the aDAE and sDAE for the example shown in Section \ref{sec:exp}. \textcolor{red}{However, we will show the performance difference between deep aDAE and deep sDAE later in Section \ref{sec:results}. The default size of the bottleneck code is 50 and 200 for aDAE and sDAE, respectively. For sDAE, there is a trade-off when setting the bottleneck code size, to avoid the high input dimension for the back-end DNN classifier, as well as for reconstructing the multiple-frame output. Typically, the weights between the encoder and the decoder are tied. Here we set them to be untied to retain more contextual information in the bottleneck codes. More specifically, the input frame number in the DAE input layer is chosen as seven for the reason that 91-frame expansion will be used in the back-end DNN classifier. In addition, larger frame expansion in DAE is more difficult to encode into a fixed bottleneck code.}

As the output of DAE is a real-valued feature, MSE was adopted as the objective function to fine-tune the whole deep DAE model. A stochastic gradient descent algorithm is performed in mini-batches with multiple epochs to improve learning convergence as follows,
\begin{equation}
Er=\frac{1}{N}\sum_{n=1}^{N}\|\hat{\textbf{X}}_{n} (\textbf{X}{_{n-\tau}^{n+\tau}},\textbf{W}, \textbf{b})-\textbf{X}_{n}\|_{2}^2
\label{DNNerrors}
\end{equation}
where $Er$ is the mean squared error, $\hat{\textbf{X}}_{n}(\textbf{X}{_{n-\tau}^{n+\tau}},\textbf{W},\textbf{b})$ and $\textbf{X}_{n}$ denote the reconstructed and input feature vector at sample index $n$, respectively, with $N$ representing the mini-batch size, $ \textbf{X}{_{n-\tau}^{n+\tau}} $ being the input audio feature vector where the window size is $2\tau+1$. $(\textbf{W}, \textbf{b})$ denote the weight and bias parameters to be learned.

The activation function of the bottleneck layer is another key point in the proposed deep DAE based unsupervised feature learning framework. Sigmoid is not suitable to be used as the activation function of the code layer, since it compresses the value of the new feature into a range [0, 1] which will reduce its representation capability. Hence, Linear or ReLU activation function is a more suitable choice. In \cite{hinton2006reducing}, the activation function of the units of the bottleneck layer or the code layer of the deep DAE is linear. A perfect reconstruction of the image can be obtained. In this work, ReLU and Linear activation functions of the bottleneck layer are both verified to reconstruct the audio features in the deep auto-encoder framework.
%Fig. \ref{fig:cv_linear_relu_ser} shows the mean squared error over the CV set with the Linear or ReLU as the activation function of the bottleneck layer units from epoch 20 to epoch 50. It can be found that the ReLU can be slightly better than the Linear function at the final epoch. Hence, the ReLU was used as the activation function of the bottleneck layer.
Note that all of the other layer units also adopt ReLU as the activation function.

\textcolor{red}{In summary, the new feature derived from the bottleneck layer of the deep auto-encoder can be regarded as the optimized feature due to three factors. The first one is that the DAE learned feature is generated from contextual input frames with new compact representations. This kind of features are better for capturing the temporal structure information compared with the original feature. The second advantage is that the deep DAE based unsupervised feature learning can smooth the disordered background noise in the audio recordings to alleviate the mismatch problem between the training set and test set. Finally, with this framework, the large amount of unlabeled data could be utilized and more statistical knowledge in the feature space can be learned.}

\section{data description and experimental setup}
\label{sec:exp}
\subsection{DCASE2016 data set for audio tagging}
\label{ssec:data_set}
The data that we used for evaluation is the dataset of Task 4 of the DCASE 2016 \cite{dcase_t4}, which is built based on the CHiME-home dataset \cite{foster2015}.
The audio recordings were made in a domestic environment \cite{christensen2010}.
Prominent sound sources in the acoustic environment are 
two adults and two children, television and electronic gadgets, 
kitchen appliances, footsteps and knocks produced by human activity, 
in addition to sound originating from outside the house \cite{christensen2010}. 
The audio data are provided as 4-second chunks at two sampling rates 
(48kHz and 16kHz) with the 48kHz data in stereo and the 16kHz data in mono.
The 16kHz recordings were obtained by downsampling the right channel of the 48kHz recordings. Note that Task 4 of the DCASE 2016 challenge is based on using only 16kHz recordings.

For each chunk, multi-label annotations were first obtained from each of 3 annotators. There are 4378 such chunks available, referred to as \textit{CHiME-Home-raw} \cite{foster2015}; discrepancies between annotators are resolved by conducting a majority vote for each label.
The annotations are based on a set of 7 label classes as shown in Table \ref{tab:annotations}.
A detailed description of the annotation procedure is provided in \cite{foster2015}.
\begin{table}[h]
	\centering
	\caption{Labels used in annotations.}
	\begin{tabular}{|c|c|}
		\hline
		Label & Description \\ \hline
		b & Broadband noise \\ 
		c & Child speech \\
		f & Adult female speech \\
		m & Adult male speech \\
		o & Other identifiable sounds \\
		p & Percussive sounds, \\
		& e.g. crash, bang, knock, footsteps \\
		v & Video game/TV \\
		\hline
	\end{tabular}
	\label{tab:annotations}
\end{table}
To reduce uncertainty about annotations, evaluations are based on considering only those chunks where 2 or more annotators agreed about label presence across label classes. There are 1946 such chunks available, referred to as \textit{CHiME-Home-refined} \cite{foster2015}. Another 816 refined chunks are kept for the final evaluation set of Task 4 of the DCASE 2016 challenge.

\begin{table}[h]
\centering
\caption{The number of audio chunks for training and test for the development set and the final evaluation set.}
\begin{tabular}{|c|c|c|}
\hline
 Fold index & \#Training & \#Test \\ \hline
 0 & 4004 & 383 \\ 
 1 & 3945 & 442 \\
 2 & 3942 & 463 \\
 3 & 4116 & 271 \\
 4 & 4000 & 387 \\
 \hline
 \hline
 \hline
 Evaluation set & 4387 & 816 \\
\hline
\end{tabular}
\label{tab:eval_data}
\end{table}

%\begin{table*}[t]
%	\centering
%	\caption{EER comparisons on seven labels among the proposed aDAE-DNN, sDAE-DNN, DNN baseline trained on MBK, DNN baseline trained on MFCC methods, SVM trained on chunks, SVM trained on frames and GMM methods, which are evaluated across the five standard folds in the development set of the DCASE 2016 audio tagging challenge.}
%	% Table generated by Excel2LaTeX from sheet 'Sheet1'
%	\begin{tabular}{l|c|c|c|c|c|c|c|c}
%		\hline
%		Development set & b  & c  & f  & m  & o  & p  & v  & Average \\
%		\hline
%		GMM (DCASE baseline)  & 0.0739 & 0.2246 & 0.2895 & 0.2692 & 0.2896 & 0.2479 & 0.0504 & 0.2065 \\
%		Chunk-SVM & 0.2121 & 0.2037 & 0.4150 & 0.4148 & 0.4104 & 0.3168 & 0.0572 & 0.2900 \\
%		Frame-SVM & 0.4800 & 0.2538 & 0.3378 & 0.4149 & 0.4395 & 0.3458 & 0.1125 & 0.3406 \\
%		MFCC-DNN & 0.0776 & 0.1450 & 0.2297 & 0.1257 & 0.2682 & 0.1829 & 0.0287 & 0.1511 \\
%		MBK-DNN-baseline & \textbf{0.0671} & 0.1418 & {0.2063} & 0.1024 & 0.2560 & 0.1485 & 0.0255 & 0.1354 \\
%		Proposed sDAE-DNN & 0.0677 & 0.1342 & {0.2061} & \textbf{0.0874} & 0.2381 & 0.1461 & \textbf{0.0228} & 0.1289 \\
%		Proposed aDAE-DNN & \textbf{0.0671} & \textbf{0.1243} & \textbf{0.2016} & 0.0923 & \textbf{0.2313} & \textbf{0.1425} & 0.0234 & \textbf{0.1261} \\
%		\hline
%	\end{tabular}%
%	\label{tab:overall_dev}
%\end{table*}


\subsection{Experimental setup}\label{ssec:exp_setup}

In our experiments, following the original specification of
Task 4 of the DCASE 2016 \cite{dcase_t4},
we use the same five folds from the given development dataset, and 
use the remaining audio recordings for training.
Table \ref{tab:eval_data} lists 
the number of chunks of training and test data used for each fold and also the final evaluation setup.

To keep the same feature configurations as in the DCASE 2016 baseline system, we pre-process each audio chunk by 
segmenting them using a $20$ms sliding window with
a $10$ms hop size, and converting
each segment into 24-dimension MFCCs and 40-dimension MBKs. For each 4-second chunk, 399 frames of MFCCs are obtained. A large set of frames expansion is used as the input of the DNN. The impact of the number of frame expansion on the performance will be evaluated in the following experiments.
%A 91-frame expansion as the input instead of the total frames were found to be better because this relaxed input scheme can increase the total number of training samples.
Hence the input size of DNN was the number of expanded frames plus the appended background noise vector. All of the input features are normalized into zero-mean and unit-variance. The first hidden layer with 1000 units and the second with 500 units were used to construct a shrinking structure \cite{zhang2014improving}. The 1000 or 500 hidden units are a common choice in DNNs \cite{mohamed2012acoustic}. Seven sigmoid outputs were adopted to predict the seven tags. The learning rate was 0.005. The momentum was set to be 0.9. The dropout rates for input layer and hidden layer were 0.1 and 0.2, respectively. The mini-batch size was 100. $T$ in Equation (\ref{eq:nat}) was 6. In addition to the \textit{CHiME-Home-refined} set \cite{foster2015} with 1946 chunks, the remaining 2432 chunks in the \textit{CHiME-Home-raw} set \cite{foster2015} without `strong agreement' labels in the development dataset were also added into the DNN training considering that DNN has a better fault-tolerant capability. Meanwhile, these 2432 chunks without `strong agreement' labels were also added into the training data for GMM and SVM training. The deep aDAE or deep aDAE has 5 layers with 3 hidden layers. For aDAE, the input is 7-frame MBKs, and the output is the middle frame MBK. The first and third hidden layer both have 500 hidden units while the middle layer is the bottleneck layer with 50 units. For sDAE, the output is 7-frame MBKs, and the middle layer is the bottleneck layer with 200 units. The dropout level for the aDAE or sDAE is set to be 0.1. The final DAE models are trained at epoch 100.

For performance evaluation, we use equal error rate (EER) as the main metric which is also suggested by the DCASE 2016 audio tagging challenge.
EER is defined as the point of the graph of false negative ($FN$) rate versus false
positive ($FP$) rate \cite{murphy2012}. The number of true positives is denoted as $TP$. EERs are computed individually for each evaluation fold,
and we then average the obtained EERs across the five folds
to get the final performance. Precision, Recall and F-score are also adopted to evaluate the performance among different systems.

%$~~~~~~~~~~~~~~~~FNR = \dfrac{\#false ~negative}{ \#positive}$\\
%\vspace{2mm}\\
%$~~~~~~~~~~~~~~~~~~~~FPR = \dfrac{\#false ~positive}{\#negative}$\\

\begin{equation}
Precision=\frac{TP}{TP+FP}
\end{equation}
\begin{equation}
Recall=\frac{TP}{TP+FN}
\end{equation}
\begin{equation}
F-score=\frac{2Precision{\cdot}Recall}{Precision+Recall}
\end{equation}

All the source codes for this paper and pre-trained models can be downloaded at Github website\footnote{\url{https://github.com/yongxuUSTC/aDAE_DNN_audio_tagging}}. The codes for the SVM and GMM baselines are also uploaded at the same website.

\subsection{Compared methods}
For a comparison, we also ran two baselines
using GMMs and the SVMs mentioned
in the Appendix section. For the GMM-based method, the number of mixture components is 8 which is a default configuration of the DCASE 2016 challenge. 
%Since the GMM based baseline focuses on computing frame-level
%likelihoods and MI-SVM prefers to instance-level scores,
The sliding window and hop size set for the two baselines and our proposed methods
are all the same. Additionally, we also use chunk-level features to evaluate on SVM-based method according to \cite{mandel2005song}. The mean and covariance of the MFCCs over the duration of the chunk can describe the Gaussian with the maximum likelihood \cite{mandel2005song}. Hence those statistics can also be unwrapped into a vector as a chunk-level feature to train the SVM. %The GMM based baseline uses Task 4 with 10ms hop size,
%while the sliding window and hop size for MI-SVM are set to be 400ms and 200ms, respectively. 
To handle audio tagging with SVM, each audio recording 
will be viewed as a bag. 
To accelerate computation, we use linear kernel function in our experiments.

We also compared our methods with the state-of-the-art methods. Lidy-CQT-CNN \cite{lidy2016cqt}, Cakir-MFCC-CNN \cite{cakirdomestic} and Yun-MFCC-GMM \cite{yundiscriminative} are the first, second and third prize of the audio tagging task of the DCASE2016 challenge \cite{dcase_t4}. The former two methods used convolutional neural networks (CNN) as the classifier. Yun-MFCC-GMM \cite{yundiscriminative} adopted the discriminative training method on GMMs.

\section{Results and discussions}
\label{sec:results}

In this section, the overall evaluations on the development set and the evaluation set of the DCASE 2016 audio tagging task will be firstly presented. Then several evaluations on the parameters of the models will be given.
\begin{table*}[t]
	\centering
	\caption{EER comparisons on seven labels among the proposed aDAE-DNN, sDAE-DNN, DNN baseline trained on MBK, DNN baseline trained on MFCC methods, Yun-MFCC-GMM \cite{yundiscriminative}, Cakir-MFCC-CNN \cite{cakirdomestic}, Lidy-CQT-CNN \cite{lidy2016cqt}, SVM trained on chunks, SVM trained on frames and GMM methods \cite{dcase_t4}, which are evaluated on the development set and the evaluation set of the DCASE 2016 audio tagging challenge.}
	\begin{tabular}{|l|c|c|c|c|c|c|c|c|}
		\hline
		Tags & b   & c   & f   & m   & o   & p   & v   & Average \\
		\hline
		& \multicolumn{8}{c|}{Development Set} \\
		\hline
		GMM (DCASE baseline) \cite{dcase_t4} & 0.074 & 0.225 & 0.289 & 0.269 & 0.290 & 0.248 & 0.050 & 0.206 \\
		Chunk-SVM & 0.464 & 0.438 & 0.430 & 0.470 & 0.524 & 0.518 & 0.274 & 0.445 \\
		Frame-SVM & 0.205 & 0.199 & 0.284 & 0.390 & 0.361 & 0.308 & 0.090 & 0.263 \\
		Yun-MFCC-GMM \cite{yundiscriminative} & 0.074 & 0.165 & 0.249 & 0.216 & 0.278 & 0.210 & 0.039 & 0.176 \\
		Cakir-MFCC-CNN \cite{cakirdomestic} & 0.070 & 0.210 & 0.250 & 0.150 & 0.260 & 0.210 & 0.050 & 0.171 \\
		Lidy-CQT-CNN * \cite{lidy2016cqt} & -   & -   & -   & -   & -   & -   & -   & - \\
		MFCC-DNN & 0.078 & 0.145 & 0.230 & 0.126 & 0.268 & 0.183 & 0.029 & 0.151 \\
		MBK-DNN & \textbf{0.067} & 0.142 & 0.206 & 0.102 & 0.256 & 0.148 & 0.025 & 0.135 \\
		Proposed sDAE-DNN & 0.068 & 0.134 & 0.206 & \textbf{0.087} & 0.238 & 0.146 & \textbf{0.023} & 0.129 \\
		Proposed aDAE-DNN & \textbf{0.067} & \textbf{0.124} & \textbf{0.202} & 0.092 & \textbf{0.231} & \textbf{0.143} & 0.023 & \textbf{0.126} \\
		\hline
		\hline
		& \multicolumn{8}{c|}{Evaluation Set} \\
		\hline
		GMM (DCASE baseline) \cite{dcase_t4} & 0.117 & 0.191 & 0.314 & 0.326 & 0.249 & 0.212 & 0.056 & 0.209 \\
		Chunk-SVM & 0.032 & 0.385 &0.407 & 0.472  & 0.536  & 0.506  & 0.473  & 0.402 \\
		Frame-SVM & 0.129 & 0.166 &0.241 & 0.353  & 0.336  & 0.268  & 0.093 & 0.227 \\
		Yun-MFCC-GMM \cite{yundiscriminative} & 0.032 & \textbf{0.177} & \textbf{0.179} & 0.253 & 0.266 & 0.207 & 0.102 & 0.174 \\
		Cakir-MFCC-CNN \cite{cakirdomestic} & 0.022 & 0.25 & 0.25 & 0.159 & 0.258 & 0.208 & 0.027 & 0.168 \\
		Lidy-CQT-CNN \cite{lidy2016cqt} & 0.032 & 0.21 & 0.214 & 0.182 & 0.32 & 0.168 & 0.035 & 0.166 \\
		MFCC-DNN & 0.032 & 0.204 & 0.21 & 0.209 & 0.288 & 0.194 & 0.039 & 0.168 \\
		MBK-DNN & 0.032 & 0.184 & 0.204 & 0.172 & 0.272 & 0.179 & 0.053 & 0.157 \\
		Proposed sDAE-DNN & 0.023 & 0.184 & 0.203 & 0.165 & 0.280 & \textbf{0.174} & 0.041 & 0.153 \\
		Proposed aDAE-DNN & \textbf{0.014} & 0.210 & 0.207 & \textbf{0.149} & \textbf{0.256} & 0.175 & \textbf{0.022} & \textbf{0.148} \\
		\hline
		\multicolumn{8}{l}{* Lidy-CQT-CNN \cite{lidy2016cqt} did not measure the EER results on the development set.}
	\end{tabular}%

	\label{tab:overall_eval}
\end{table*}

\begin{table*}[t]
	\centering
	\caption{Precision, Recall and score comparisons between the MBK-DNN baseline and the aDAE-DNN method, which are evaluated for seven tags on the final evaluations set of the DCASE2016 audio tagging task.}
	\begin{tabular}{c|c|c|c|c|c|c|c|c|c}
		\hline
		\multicolumn{2}{c|}{Evaluation set} & b  & c  & f  & m  & o  & p  & v  & Average \\
		\hline
		\multirow{2}[0]{*}{Precision} & MBK-DNN & 1.000 & \textbf{0.684} & 0.676 & 0.464 & 0.583 & 0.774 & 0.980 & 0.737 \\
		& aDAE-DNN & 1.000 & 0.654 & \textbf{0.691} & \textbf{0.474} & \textbf{0.692} & \textbf{0.819} & \textbf{0.991} & \textbf{0.760} \\
		\hline
		\multirow{2}[0]{*}{Recall} & MBK-DNN & 0.968 & 0.905 & \textbf{0.507} & 0.405 & 0.224 & \textbf{0.677} & \textbf{0.976} & 0.666 \\
		& aDAE-DNN & 0.968 & \textbf{0.912} & 0.464 & \textbf{0.456} & \textbf{0.288} & 0.658 & 0.973 & \textbf{0.674} \\
		\hline
		\multirow{2}[0]{*}{F-score} & MBK-DNN & 0.984 & \textbf{0.780} & \textbf{0.580} & 0.432 & 0.324 & 0.722 & 0.978 & 0.686 \\
		& aDAE-DNN & 0.984 & 0.762 & 0.556 & \textbf{0.465} & \textbf{0.407} & \textbf{0.730} & \textbf{0.982} & \textbf{0.698} \\
		\hline
	\end{tabular}%
	\label{tab:prec_rec_f1_eval}
\end{table*}

\subsection{Overall evaluations}
%\begin{figure}[t]
%	\centering
%	\includegraphics[width=\columnwidth]{DNNAE_DNN_GMM_MISVM.pdf} 
%	\caption{Equal error rates obtained using the proposed fully DNN approach, the fully DNN improved by the proposed deep auto-encoder feature (denoted as DNN\_AE), and the two baselines, namely GMM and MISVM across five evaluation folds.}
%	\label{fig:dnn_gmm_misvm}
%\end{figure}

\textcolor{red}{Table \ref{tab:overall_eval} shows the EER comparisons on seven labels among the proposed aDAE-DNN, sDAE-DNN, DNN baseline trained on MBK, DNN baseline trained on MFCC methods, Yun-MFCC-GMM \cite{yundiscriminative}, Cakir-MFCC-CNN \cite{cakirdomestic}, Lidy-CQT-CNN \cite{lidy2016cqt}, SVM trained on chunks, SVM trained on frames and GMM methods \cite{dcase_t4}, which are evaluated on the development set and the evaluation set of the DCASE 2016 audio tagging challenge.
On the development set, it is clear that the proposed DNN-based approaches outperform
the SVM and GMM baselines across the five-fold evaluations. GMM is better than the SVM methods. SVM performs worse on the audio event `b' where less training samples are included in the imbalanced development set compared with other audio events \cite{dcase_t4}. However, the GMM and DNN methods perform better on the audio event `b' with lower EERs. The chunk-level SVM is inferior to the frame-level SVM. This is because the audio tagging is a multi-label classification task not single-label classification task while the statistical mean value in the chunk-SVM will make the feature indistinct among different labels in the same chunk. Compared with the DNN methods, SVM and GMM are less effective in utilizing the contextual information and the potential relationship among different tags. Note that the DNN models here are all trained using binary cross-entropy defined in Eq. (\ref{eq:DNNerrors_bce}) as the loss function. Binary cross-entropy is found better than the mean squared error for training the audio tagging models which will be shown in the following subsection. The DNN trained on MFCCs is worse than the DNN trained on MBKs with the reduced EER from 0.151 to 0.135, especially on the percussive sounds (`p'), e.g. crash, bang, knock and footsteps. This result is consistent with the observations in speech recognition using DNNs \cite{seltzer2013investigation}. Compared with MBKs, MFCCs lost some information after the discrete cosine transformation (DCT) step. The bottleneck code size for aDAE-DNN and sDAE-DNN here is 50 and 200, respectively. It is found that aDAE-DNN can reduce the EER from 0.157 to 0.148 compared with MBK-DNN-baseline, especially on tag `c' and tag `o'. The aDAE-DNN method is slightly better than the sDAE-DNN because the sDAE-DNN should have a larger bottleneck code to reconstruct the seven-frame output. However, the large size of the bottleneck code in sDAE-DNN will make the input dimension of the back-end DNN classifier very high.} \textcolor{red}{Lidy-CQT-CNN \cite{lidy2016cqt} did not measure the EER on the development set \cite{lidy2016cqt}. Our proposed DNN methods can get better performance than Cakir-MFCC-CNN \cite{cakirdomestic} and Yun-MFCC-GMM \cite{yundiscriminative}.}
%\footnote{\url{https://uk.mathworks.com/help/stats/ttest.html}, MATLAB verssion is R2016a}.

\begin{figure}[t]
	\centering
	\centerline{\includegraphics[width=\columnwidth]{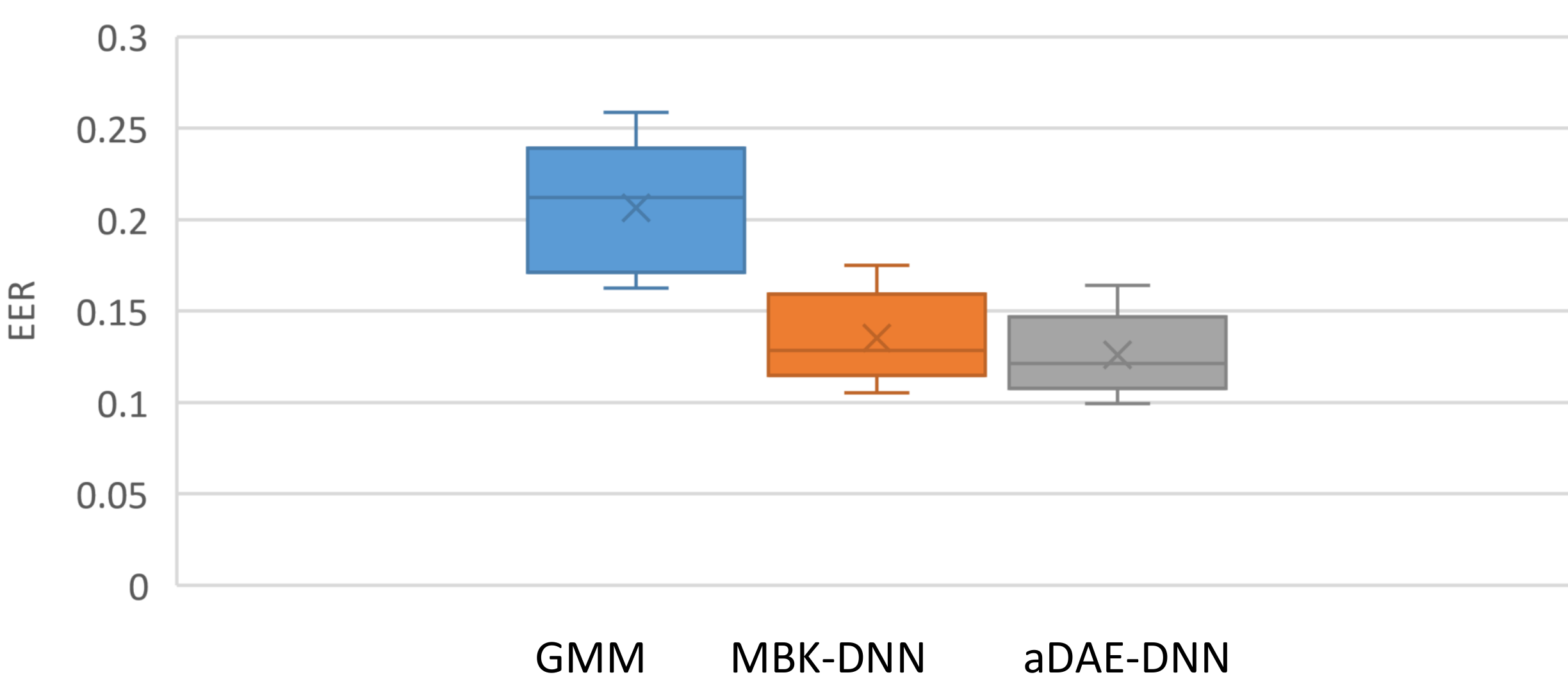}}
	\caption{The box-and-whisker plot of EERs, among the GMM baseline, Mel-Filter bank (MBK)-DNN baseline and asymmetric DAE (aDAE)-DNN method, across five standard folds on the development set of the DCASE 2016 audio tagging challenge.}
	\label{fig:box_whisker_plot_eer}
\end{figure}

\textcolor{red}{Fig. \ref{fig:box_whisker_plot_eer} shows the box-and-whisker plot of EERs, among the GMM baseline, MBK-DNN baseline and aDAE-DNN method, across five standard folds on the development set of the DCASE 2016 audio tagging challenge. It can be found that the aDAE-DNN is consistently better than the MBK-DNN baseline. To test the statistical significance between the aDAE-DNN and MBK-DNN baseline, we use the paired-sample \textit{t}-test tool in MATLAB. The audio tagging task of the DCASE2016 challenge has five standard folds with seven tags. Hence, a 35-dimension vector can be obtained for each method, then the paired-sample \textit{t}-test tool can be used to calculate the \textit{p}-value. Its results indicate that \textit{t}-test rejects the null hypothesis at the 1\% significance level. It was found that the \textit{p}-value is $\ll 0.01$ in this test which indicates that the improvement is statistically significant. Finally our proposed method can get a 38.9\% relative EER reduction compared with the GMM baseline of the DCASE 2016 audio tagging challenge on the development set.}
%This is because of the following two main reasons:
%First, our proposed approach can well utilize
%the long-term temporal information instead of treating
%those information independently. Second, it can
%map the whole audio features sequence
%into a multi-tag vector by working
%as an encoding function. However, GMM and SVM based methods build the models only on single instances. The contextual information and the potential relationship among different tags were not well utilized.

\begin{figure}[t]
	\centering
	\centerline{\includegraphics[width=\columnwidth]{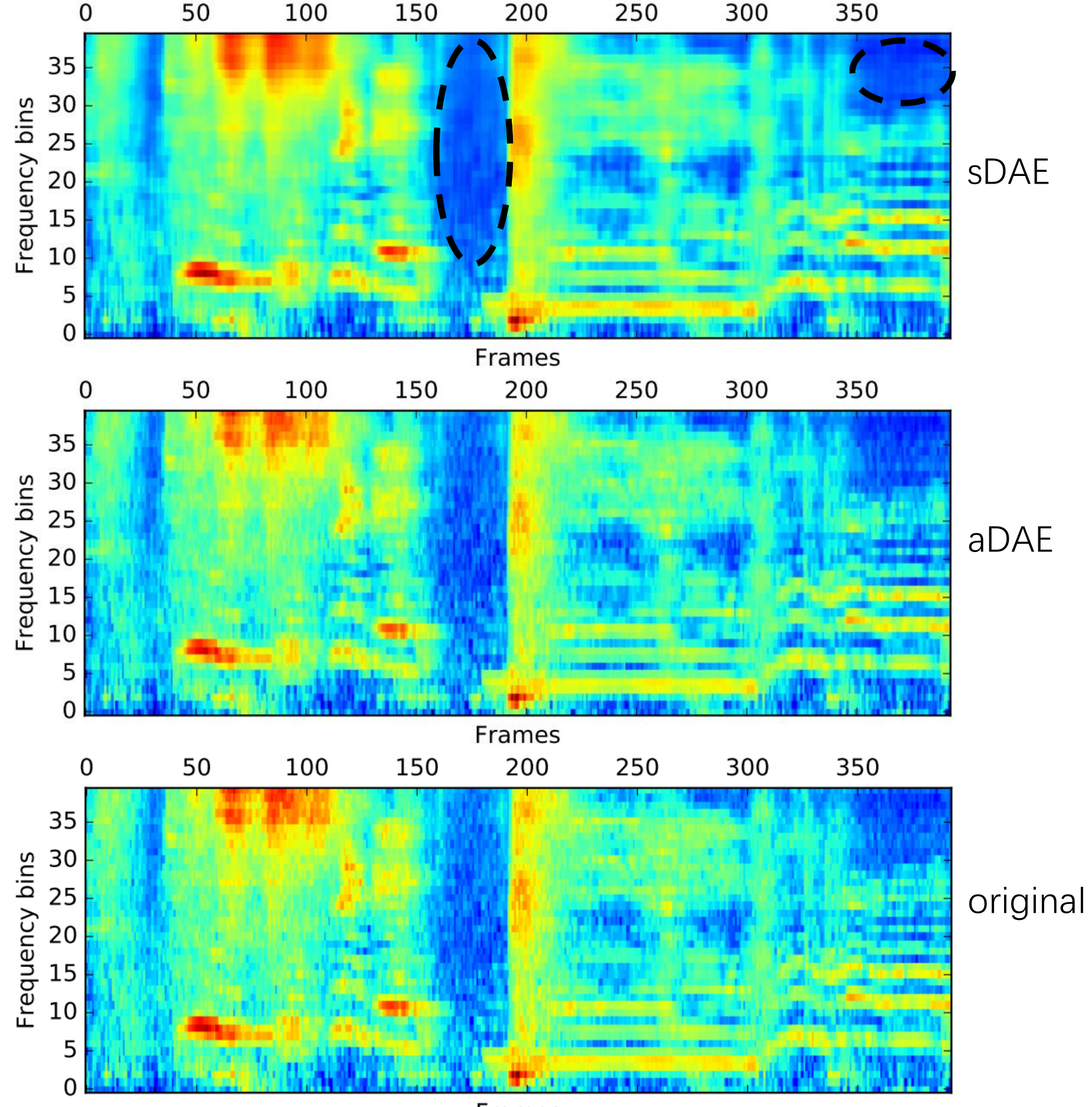}}
	\caption{Spectrograms of the reconstructed Mel-Filter Banks (MBKs) by the deep asymmetric DAE (aDAE) and deep symmetric DAE (sDAE), and also the original MBKs. The dotted ovals indicate the smoothed parts on the reconstructed MBKs.}
	\label{fig:reconstructedMFCC}
\end{figure}

\textcolor{red}{Table \ref{tab:overall_eval} also presents EER comparisons on the evaluation set. Note that the final evaluation set was not used for any training which means sDAE and aDAE also did not use it in the training. It can be found that our proposed aDAE-DNN can get the state-of-the-art performance. Our MBK-DNN is a strong baseline through the use of several techniques, e.g., the dropout, background noise aware training, shrinking structure and also binary cross-entropy. The proposed aDAE-DNN can get a 5.7\% relative improvement compared with the MBK-DNN baseline. sDAE-DNN did not show improvement over the MBK-DNN because sDAE-DNN with the bottleneck code size 200 can not well reconstruct the unseen evaluation set. However the aDAE-DNN with the bottleneck code size 50 can well reconstruct the unseen evaluation set. Finally, our proposed methods can get the state-of-the-art performance with 0.148 EER on the evaluation set of the DCASE 2016 audio tagging challenge. Another interesting result here is that Yun-MFCC-GMM \cite{yundiscriminative} performs well on tag `c' and tag `f' where high pitch information exists. It would be interesting to fuse their prediction posteriors together in our future work.}

\textcolor{red}{To give a further comparison between the MBK-DNN baseline and the aDAE-DNN method, Table \ref{tab:prec_rec_f1_eval} shows precision, recall and score comparisons evaluated for seven tags on the final evaluation set of the DCASE2016 audio tagging task. As the DNN prediction belongs to [0,1], a threshold 0.4 is set to judge whether it is a hit or not. Using aDAE-DNN better performance than the MBK-DNN baseline can be obtained on most of the three measures. One interesting result is that DNN method can get a quite high score on tag `b' although there are only few training samples in the training set \cite{dcase_t4}.}

%Compared with the fully DNN using the basic MFCC feature, the deep auto-encoder feature can further reduce the EERs across all of the five folds. This is because the deep auto-encoder model can well reconstruct the MFCC features. 
\textcolor{red}{Fig. \ref{fig:reconstructedMFCC} shows the spectrograms of the original MBKs and the reconstructed MBKs by the deep sDAE and deep aDAE. Both of them can reconstruct the original MBKs well while sDAE got a smoother reconstruction. There is background noise in original MBKs which will lead to the mismatch problem mentioned earlier. sDAE can well reduce the background noise shown in the dashed ellipses with the risk of losing the important spectral information. However, aDAE can be a trade-off between background noise smoothing and signal reconstruction. On the other hand, the weights of the encoder and decoder in the deep aDAE and the deep sDAE are not typically tied. In this way, more contextual information is encoded into the bottleneck layer to get a compact representation, which is helpful for the audio tagging task considering the fact that the reference labels are in chunk-level.}

\subsection{Evaluations for the number of contextual frames in the input of the DNN classifier}
\begin{figure}[t]
	\centering
	\centerline{\includegraphics[width=\columnwidth]{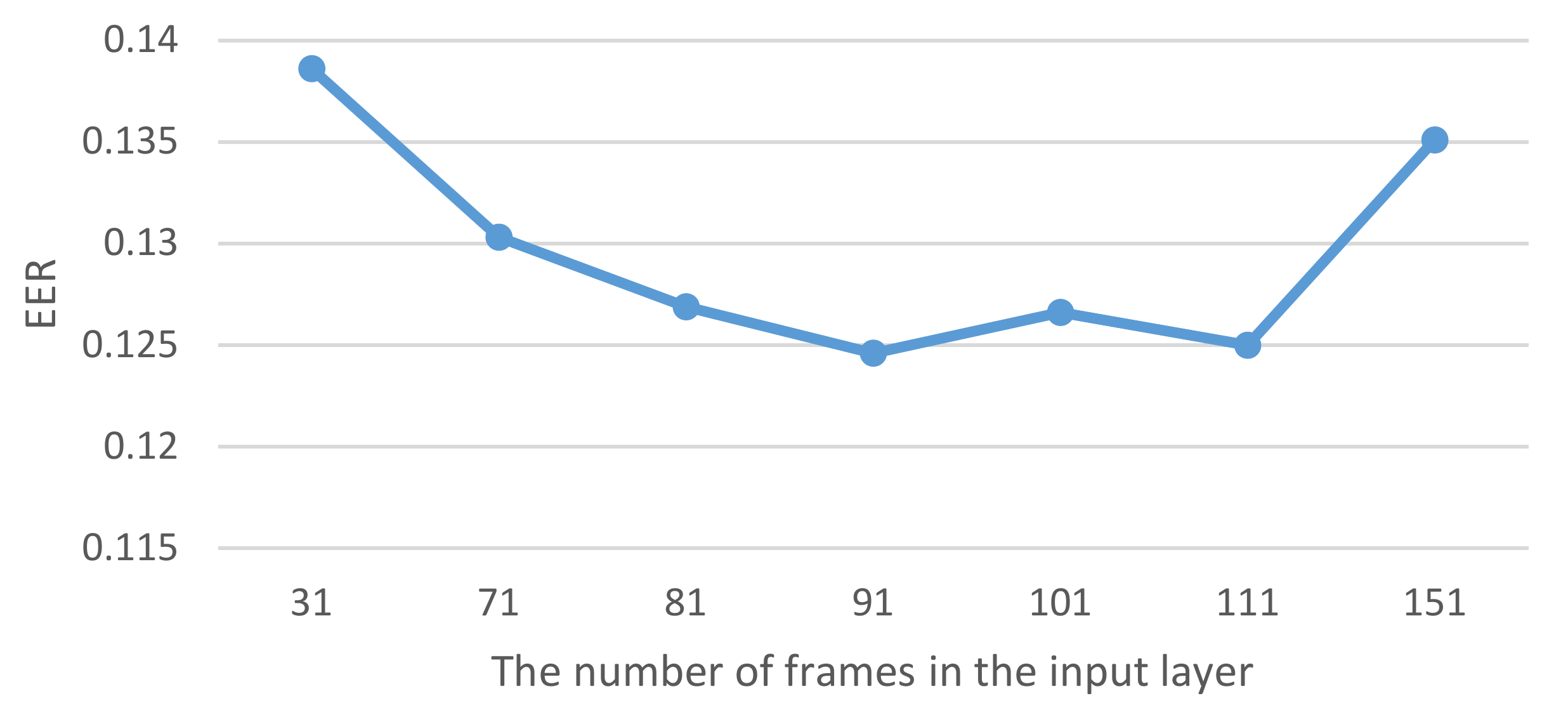}}
	\caption{EERs on Fold 1 of the development set evaluated using different number of frame expansions in the input layer of DNNs.}
	\label{fig:different_frames}
\end{figure}
The reference label information for this audio tagging task is on the utterance-level rather than the frame-level, and the occurring orders and frequencies of the tags are unknown. \textcolor{red}{Hence, it is important to use a large set of the contextual frames in the input of the DNN classifier. However, the dimension of the input layer of the DNN classifier will be too high and the number of training samples would be reduced if the number of the frame expansion is too large. Larger input size will increase the complexity of the DNN model and as a result, some information could be lost during the feed-forward process considering that the hidden unit size is fixed to be 1000 or 500. Fewer training samples will make the training process of DNN unstable considering that the parameters are updated using a stochastic gradient descent algorithm performed in mini-batches.}
%We proposed to suitably relax the number of the contextual frames to 91-frame while the total number of frames for each utterance is 99-frame. Then the frame expansion is conducted with one frame slide along the utterance. The same tag labels are assigned to each training sample belonging to the certain utterance. This will increase the number of training samples and also ensure that most of the tags related audio features are covered in the input.

\textcolor{red}{Fig. \ref{fig:different_frames} shows the EERs for Fold 1 evaluated by using different number of contextual frames in the input of the DNN classifier. Here the MBKs are used as the input features. It can be found that using the 91-frame MBKs as the input gives the lowest EER. As mentioned in the experimental setup, the window size of each frame is 20ms with 50\% hop size. 91-frame expansion means that the input length is about one second. The length of the whole chunk is 4 seconds. It indicates that most of the tags overlap with each other in certain chunk. Meanwhile, 91-frame expansion in the input layer of the DNN is a good trade-off among the contextual information, input size, and total training samples.}
%Using the whole utterance, namely the 99-frame MFCC, as the input leads to the worst performance due to the smallest number of training samples. An interesting phenomenon is that using the 11-frame MFCC as the input still gives comparable performance, and in this case, the window used is relatively small and may not contain any audio features corresponding to certain tags in the output. This might indicate that most of the tags overlap heavily with each other in certain utterance.

\subsection{Evaluations for different kinds of input features and different types of loss functions}
\begin{figure}[t]
	\centering
	\centerline{\includegraphics[width=\columnwidth]{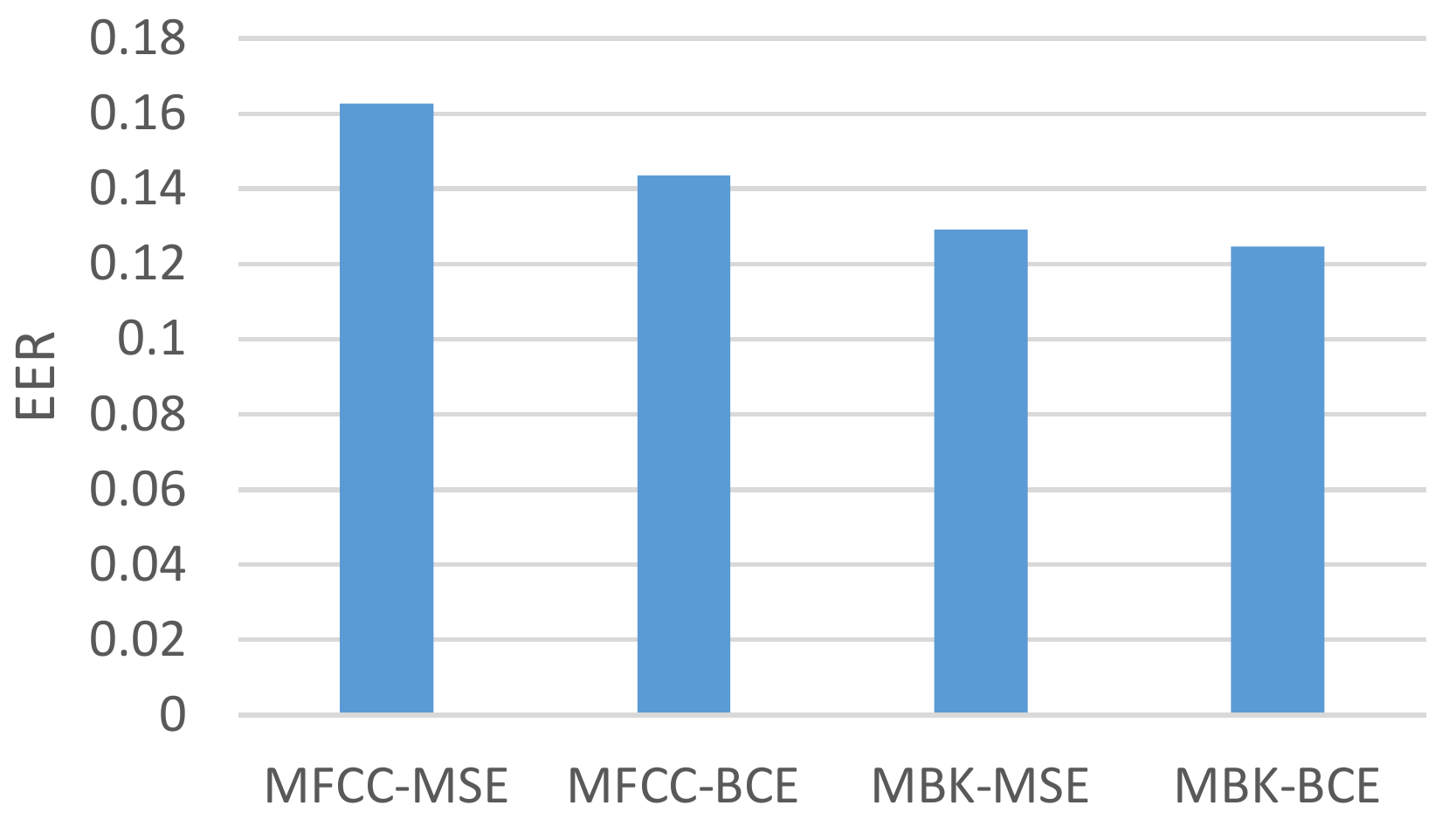}}
	\caption{EERs on Fold 1 of the development set evaluated using different features, namely MFCCs and Mel-Filter Banks (MBKs), different loss functions, namely mean squared error (MSE) and binary cross entropy (BCE).}
	\label{fig:fea_lossfunc}
\end{figure}
\textcolor{red}{Fig. \ref{fig:fea_lossfunc} shows EERs on Fold 1 evaluated using different features, namely MFCCs and MBKs, different loss functions, namely mean squared error (MSE) and binary cross entropy (BCE). It can be found that MBKs perform better than MFCCs. MBKs contain more spectral information than the MFCCs. BCE is superior to MSE considering that the value of label is binary, either zero or one. MSE is more suitable to fit the real values.}

\subsection{Evaluations for different bottleneck size of DAE and comparison with the common auto-encoder}
\begin{figure}[t]
	\centering
	\centerline{\includegraphics[width=\columnwidth]{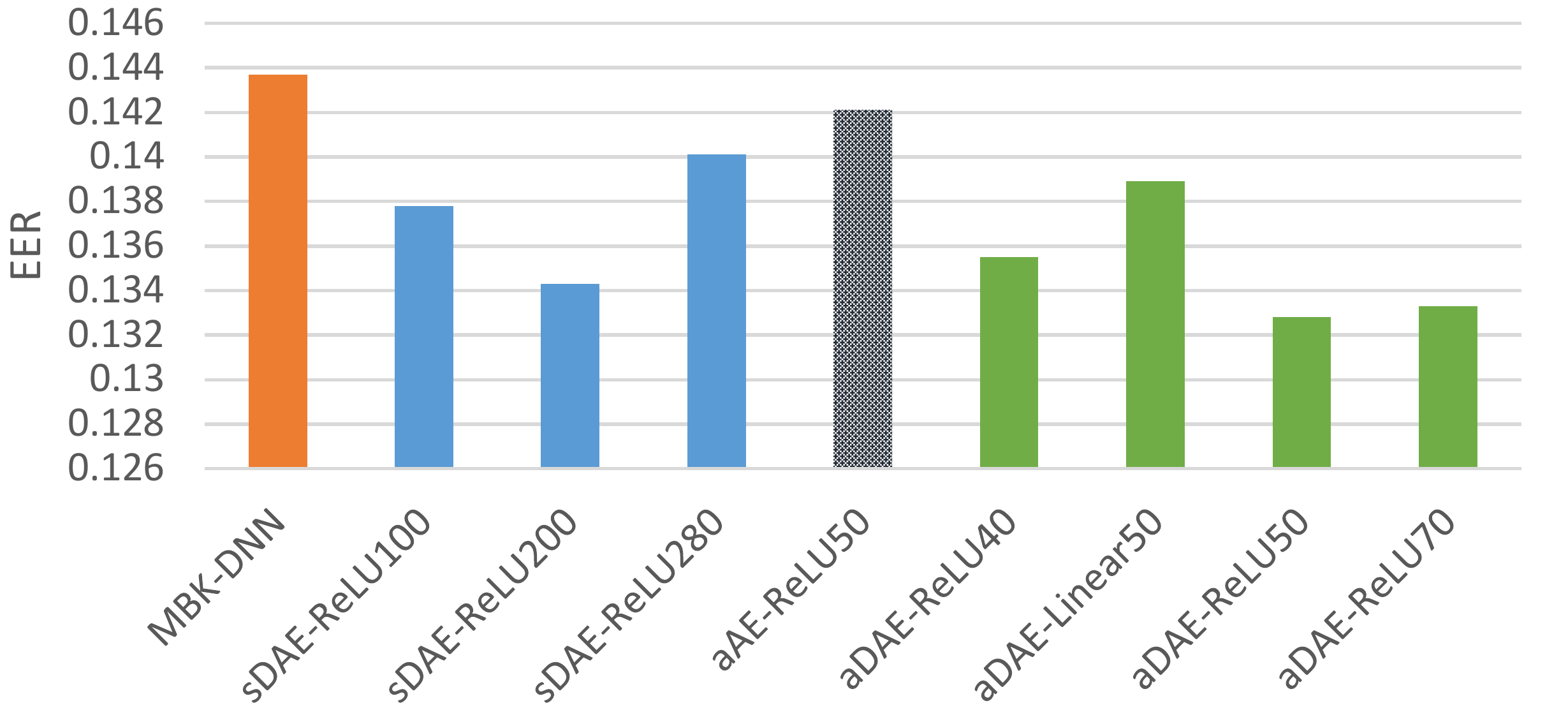}}
	\caption{EERs on Fold 0 of the development set evaluated using different de-noising auto-encoder configurations and compared with the MBK-DNN baseline. sDAE-ReLU200 means the symmetric DAE with 200 ReLU units in the bottleneck layer. aDAE-Linear50 means the asymmetric DAE with 50 linear units in the bottleneck layer. aAE-ReLU50 denotes the asymmetric auto-encoder without de-noising.}
	\label{fig:conf_dae}
\end{figure}
\textcolor{red}{Fig. \ref{fig:conf_dae} shows the EERs on Fold 0 evaluated using different de-noising auto-encoder configurations and compared with the MBK-DNN baseline. For the deep sDAE, the bottleneck layer size needs to be properly set. If it is too small, the 7-frame MBKs can not be well reconstructed. While it will increase the input size of the DNN classifier if the bottleneck code is too large. For the deep aDAE, the bottleneck layer with 50 ReLU units is found empirically to be a good choice. The linear unit (denoted as aDAE-Linear50) is worse than the ReLU unit for the new feature representation. Another interesting result is that the performance was almost the same if there is no de-noising operation (denoted as aAE-ReLU50) in the auto-encoder. The reason is that the baseline DNN is well trained on MBKs with the binary cross-entropy as the loss function.}

\subsection{Evaluations for the size of the training dataset}
%In this audio tagging task, 4378 chunks are available for system development, which are denoted as `CHiME-Home-raw' \cite{foster2015}. Among them, there are 1946 `strong agreement' chunks where two or more annotators agreed about the label presence across label classes and they are denoted as `CHiME-Home-refined' \cite{foster2015}, while the labels of the remaining chunks are not accurate. 
In the preceding experiments, `CHiME-Home-raw' dataset was used to train the DNN, GMM and SVM models. Here, to evaluate the performance using different training data sizes, DNNs were trained based on `CHiME-Home-raw' or `CHiME-Home-refined' alternatively while keeping the same testing set. MBKs were used as the input features for the DNN classifier.

Table \ref{tab:datasets} shows the EERs for Fold 1 across seven tags with the DNNs trained on the `CHiME-Home-raw' set and `CHiME-Home-refined' set. It can be clearly found that the DNN trained on the `CHiME-Home-raw' set is better than the DNN trained on the `CHiME-Home-refined' set, although part of the labels of the `CHiME-Home-raw' set are not accurate. This indicates that DNN has fault-tolerant capability which suggests that the labels for the tags can not be refined with much annotators' effort. The size of the training set is crucial for the DNN training. Nonetheless the GMM method is sensitive to the inaccurate labels. The increased training data with inaccurate tag labels does not help to improve the performance of GMMs.

\subsection{Further discussions on the deep auto-encoder features}
\begin{figure}[t]
	\centering
	\centerline{\includegraphics[width=\columnwidth]{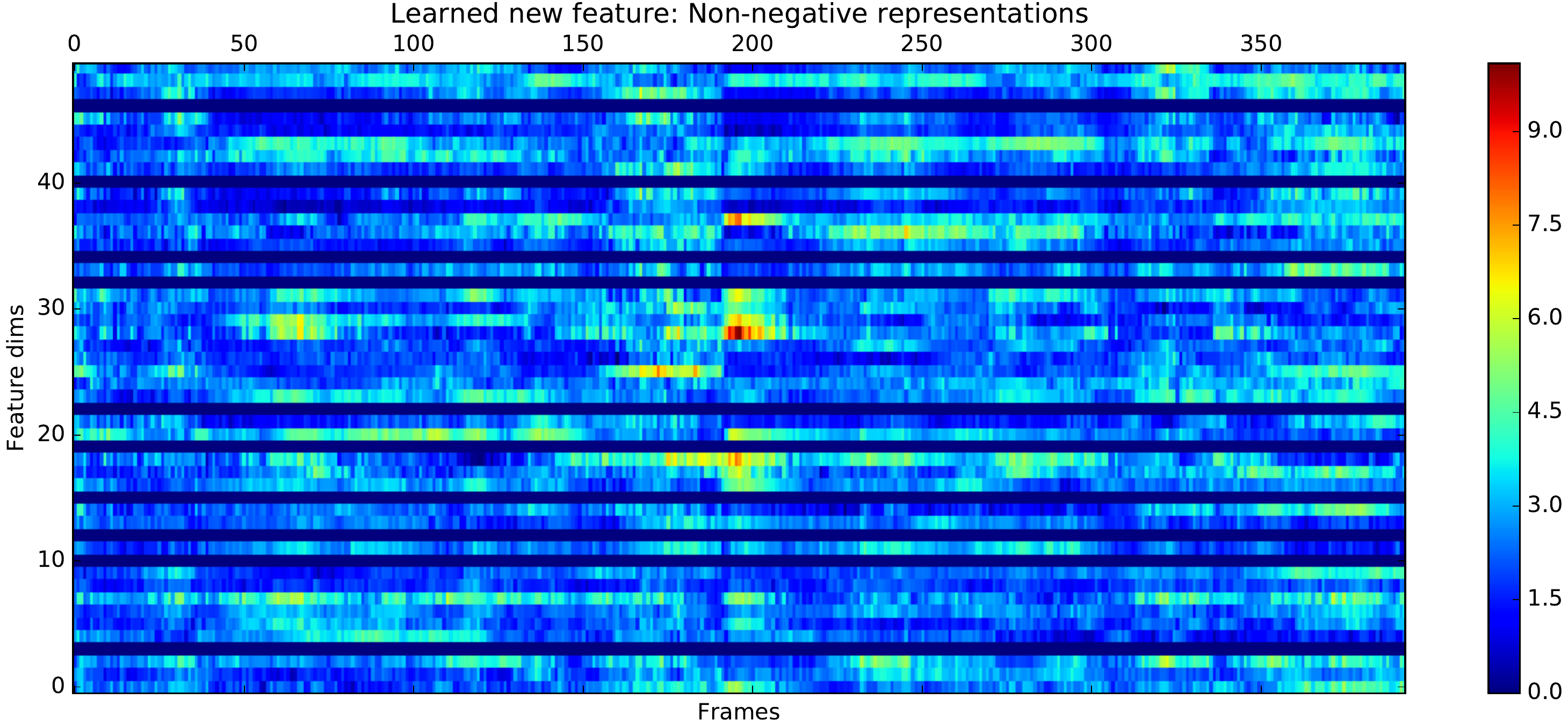}}
	\caption{The audio spectrogram of the deep asymmetric DAE (aDAE) features with the non-negative representation.}
	\label{fig:ae_fea}
\end{figure}

Fig. \ref{fig:ae_fea} presents the audio spectrogram of the deep aDAE features, which can be regarded as the new non-negative representation or optimized feature of the original MBKs. The units of the bottleneck layer in the deep aDAE are all activated by the ReLU functions as mentioned in Sec. \ref{sec:dae_FeaLearn}. Hence, the values of the learned feature are all non-negative, leading to a non-negative representation of the original MBKs. Such a non-negative representation can then be multiplied with the weights in the decoding part of the DAE to obtain the reconstructed MBKs. It is also adopted to replace the MBKs as the input to the DNN classifier to make a better prediction for the tags. The pure blue area at some dimensions in Fig. \ref{fig:ae_fea} indicates the zero values in the ReLU activation function.
	
\section{Conclusions}\label{sec:conclusions}

\textcolor{red}{In this paper we have studied the acoustic modeling and feature learning issues in audio tagging. We have proposed a DNN incorporating unsupervised feature learning based
approach to handle audio tagging with weak labels, in the sense that only the chunk-level instead of the frame-level labels are available. The Dropout and background noise aware training methods were adopted to improve its generalization capacity for new recordings in unseen environments.
%This DNN is regarded as an encoding function to map the sequence of audio features to a multi-tag vector.
A deep asymmetric DAE with untied weights based unsupervised feature learning was also proposed to generate a new feature with non-negative representations. The DAE can generate smoothed feature against the disordered background noise and also give a compact representation of the contextual frames. 
We tested our approach on the dataset of the Task 4 of the DCASE 2016 challenge,
and obtained significant improvements over the two baselines, namely GMM and SVM. Compared with the official GMM-based baseline system given in the DCASE 2016 challenge, the proposed DNN system can reduce the EER from 0.207 to 0.126 on the development set. The proposed unsupervised feature learning method can get a relative 6.7\% EER reduction compared with the strong DNN baseline on the development set. We also get the state-of-the-art performance with 0.148 EER on the evaluation set compared with the latest results (Yun-MFCC-GMM \cite{yundiscriminative}, Cakir-MFCC-CNN \cite{cakirdomestic}, Lidy-CQT-CNN \cite{lidy2016cqt}) from the DCASE 2016 challenge. For the future work, we will use convolutional neural network (CNN) to extract more robust high-level features for the audio tagging task. Larger dataset, such as Yahoo Flickr Creative Commons 100 Million (YFCC100m) dataset \cite{thomee2016yfcc100m} and YouTube-8M dataset \cite{abu2016youtube} will be considered to further evaluate the proposed algorithms.}
% Estimating the event location along the chunk based only on the weakly labeled data will also be investigated.

\appendix \label{sec:baseline_systems}
Two baseline methods compared in our work are briefly summarized below.
\subsection{Audio Tagging using Gaussian Mixture Models}\label{subsec:GMM}
GMMs are a commonly used generative
classifier. A GMM is parametrized by 
$\Theta=\{ \omega_m, \mu_m, \Sigma_m \}, m = \{1,\cdots ,M\}$,
where $M$ is the number of mixtures and $w_m$ is the weight of the $m$-th
mixture component. 

To implement multi-label classification with simple event tags, 
a binary classifier
is built associating with each audio event class in the training step. 
For a specific event class, all audio frames in an audio chunk labeled with this event
are categorized into a positive class, whereas the remaining features are categorized into 
a negative class. On the classification stage, given an audio
chunk $C_i$, the likelihoods of each audio frame
$x_{ij}, (j \in \{1 \cdots L_{C_i}\})$ are calculated for the two class models, respectively.
%
%\begin{equation}
%f(x_{ij}, \Theta_{pos}) = \sum_m w_m^{pos} N(x_{ij},~ \mu_m^{pos}, \Sigma_m^{pos})
%\end{equation}
%\begin{equation}
%f(x_{ij}, \Theta_{neg}) = \sum_m w_m^{neg} N(x_{ij},~ \mu_m^{neg}, \Sigma_m^{neg})
%\end{equation}
Given audio event class $k$ and chunk $C_i$, the classification score $S_{C_{ik}}$
is obtained as log-likelihood ratio:

\begin{equation}
S_{C_{ik}} = \sum_j \text{log}(f(x_{ij}, \Theta_{\text{positive}})) - \sum_j \text{log}(f(x_{ij}, \Theta_{\text{negative}}))
\end{equation}

\subsection{Audio Tagging using Multiple Instance SVM }\label{subsec:MIL}

Multiple instance learning is described in terms of bags $\textbf{B}$. 
The $j$th instance in the $i$th bag, $B_i$, is defined as $x_{ij}$ 
where $j \in I=\{1 \cdots l_i\}$, and $l_i$ is the number of instances in $B_i$.
$B_i$'s label is $Y_i \in \{ -1, 1 \}$. 
If $Y_i = -1$, then $x_{ij} = -1$ for all $j$. 
If $ Y_i = 1$, then at least
one instance $x_{ij} \in B_i$ is a positive example of the underlying concept \cite{andrew2003}.

As MI-SVM is the bag-level MIL support vector machine to maximize the bag margin,
we define the functional margin of a bag with respect to a hyper-plane as:

\begin{equation}
\gamma_i = Y_i \max_{j \in I} (\langle \textbf{w},\textbf{x}_{ij} \rangle +b) 
\end{equation}

Using the above notion, MI-SVM can be defined as:

\begin{equation}
\min_{\textbf{w},b,\xi} \dfrac{1}{2}\Vert \textbf{w}^2 \Vert + A\sum_i\xi_i
\end{equation} 
~~~~~~~~subject~~ to:$~~~~~~\forall_i: \gamma_i \geq 1-\xi_i $,~~~$\xi_i \geq 0 $

where $\textbf{w}$ is the weight vector, $b$ is bias, $\xi$ is
margin violation, and $A$ is a regularization parameter.

Classification with MI-SVM proceeds in two steps. 
In the first step, $\textbf{x}_i$ is initialized as the centroid
for every positive bag $B_i$ as follows

\begin{equation}
\overline{\textbf{x}}_i = \sum_{j \in I} \textbf{x}_{ij}/l_i 
\end{equation}

\begin{table}[t]
	\centering
	\caption{EERs for Fold 1 across seven tags using DNNs and GMMs trained on the `CHiME-Home-raw' set and `CHiME-Home-refined' set.}
	\begin{tabular}{|c|c|c|c|c|c|c|c|c|}
		\hline
		Dataset & b & c & f & m & o & p & v \\
		\hline
		DNN-Refine & 0.009 & 0.168 & 0.223 & 0.158 & \textbf{0.273} & 0.118 & 0.050\\ 
		
		DNN-Raw  & \textbf{0.002} & \textbf{0.124} & \textbf{0.209} & \textbf{0.146} & {0.277} & \textbf{0.089} & \textbf{0.025}\\
		\hline
		GMM-Refine & \textbf{0.000} & \textbf{0.203} & 0.343 & 0.303 & \textbf{0.305} & \textbf{0.333} & 0.154\\ 
		
		GMM-Raw  & {0.013} & {0.283} & \textbf{0.294} & \textbf{0.217} & {0.326} & {0.347} & \textbf{0.051}\\
		\hline		
	\end{tabular}
	\label{tab:datasets}
\end{table}

The second step is an iterative procedure in order to
optimize the parameters. 

Firstly, $\textbf{w}$ and $b$ are computed for the data set 
with positive samples $\{x_I: Y_i=1\}$.

Secondly, we compute

$~~~~~~~~~~~~~~~~~~~~~~~~f_{ij} = \langle\textbf{w}, \textbf{x}_{ij}\rangle +b$, $~~~~~\textbf{x}_{ij} \in \textbf{B}_i$ 

Thirdly, we change $\overline{\textbf{x}}_i$ by

$~~~~~~~~~~~~~~~~~~~~~~~~\overline{\textbf{x}}_i = \textbf{x}_j$

$~~~~~~~~~~~~~~~~~~~~~~~~j=\arg\max_{j \in I} f_{ij}, \forall I, Y_I=1$ 

The iteration in this step will stop when there is no change of $\overline{\textbf{x}}_i$. The optimized parameters will be used for testing.

\section*{Acknowledgments}
We gratefully acknowledge the critical comments by anonymous reviewers and associate editor of this paper.

\ifCLASSOPTIONcaptionsoff
 \newpage
\fi



%\bibliographystyle{IEEEtran}
\bibliographystyle{IEEEtran}
\bibliography{refs}

\end{document}